\documentclass[conference]{IEEEtran}
\IEEEoverridecommandlockouts
\usepackage{cite}
\usepackage{amsmath,amssymb,amsfonts}
\usepackage{algorithm}
\usepackage{algpseudocode}
\usepackage{enumitem}
\usepackage{graphicx}
\usepackage{textcomp}
\usepackage{xcolor}
\usepackage{booktabs}
\algnewcommand\algorithmicnot{\textbf{not}}
\algdef{SE}[IF]{IfNot}{EndIf}[1]{\algorithmicif\ \algorithmicnot\ #1\ \algorithmicthen}{\algorithmicend\ \algorithmicif}%

\def\BibTeX{{\rm B\kern-.05em{\sc i\kern-.025em b}\kern-.08em
    T\kern-.1667em\lower.7ex\hbox{E}\kern-.125emX}}
\begin{document}

\title{Identity and Access Management Framework for Multi-tenant Resources in Hybrid Cloud Computing}

\author{\IEEEauthorblockN{Saurabh Deochake}
\IEEEauthorblockA{\textit{Twitter, Inc.} \\
San Francisco, USA \\
sdeochake@twitter.com}
\and
\IEEEauthorblockN{Vrushali Channapattan}
\IEEEauthorblockA{\textit{Twitter, Inc.} \\
San Francisco, USA \\
vchannapattan@twitter.com}
}

\maketitle

\begin{abstract}
While more organizations have been trying to move their infrastructure to the cloud in recent years, there have been significant challenges in how identities and access are managed in a hybrid cloud setting. This paper showcases a novel identity and access management framework for shared resources in a multi-tenant hybrid cloud environment. The paper demonstrates a method to implement the ``mirror" identities of on-premise identities  in the cloud. Following the best security practices, the framework ensures that only rightful users can use their mirror identities in the cloud. Furthermore, the paper also proposes a technique in scaling the framework to accommodate large-scale enterprises. The framework exhibited in the paper provides a comprehensive and scalable solution for enterprises to implement identity and access control in their hybrid cloud infrastructure. Although the paper focuses on implementing the framework in Google Cloud Platform, it can be easily applied to any major public cloud platform.
\end{abstract}


\section{Introduction}\label{sec:intro}
With the advent of rapid growth in the public cloud computing, traditionally on-premise data center-based organizations have started their journey of using the public cloud. Public clouds deliver Infrastructure as a Service (IaaS) while offering scalability, elasticity, and efficiency that is much better than an organization running their infrastructure in their data centers. Since public cloud platforms offer vertical scalability, an organization can increase its scale using on-demand resource utilization \cite{bib1}. A benefit of having on-demand resource availability is that organizations do not have to buy expensive hardware in advance, thereby reducing their expenditure. However, in terms of identity and access management for the data, organizations may not be able to avail a strict control over their data security since they have to fully depend on what public clouds offer and where they store their data \cite{bib2}. On the other hand, private clouds are set up based on specific needs for an enterprise. Private clouds offer much better control of security and data protection since data is stored behind a restrictive network that is operated by an organization \cite{bib3}. Moreover, organizations running private cloud can also choose to use public cloud for publicly available data, giving more flexibility in how data is stored. However, this may result in increased complexity and expenditure for an organization since the organizations running private cloud have to manage their hardware and software, including access control and data protection software \cite{bib4}. For organizations desiring the best of both - public and private cloud while operating their data centers, hybrid cloud can be an attractive option. Hybrid cloud computing allows applications and features to work across the boundaries of public cloud and data centers. This provides much more flexibility in terms of how these services interact with each other, how the data is stored, and how data is processed based on the needs of enterprise software running in the on-premise data center while making use of all the benefits of the public cloud like the ability to scale over a broader geographical footprint for the business continuity and leveraging new cloud service offerings and capabilities as they become available \cite{bib5}. Although hybrid cloud computing offers a lot of benefits compared to traditional models of public or private cloud computing, building a frictionless identity and access management (IAM) framework that transparently works across the boundaries of private and public cloud infrastructure is especially an important challenge to solve. \\
While Twitter had a cost-effective on-premise data management infrastructure, there was a desire and motivation to extend these capabilities to the public cloud \cite{bib6}. Twitter’s Partly Cloudy is a project to extend data processing at Twitter from an on-premise-only model to a hybrid-cloud model on Google Cloud Platform (GCP) \cite{bib7}. As a part of this project, Twitter migrated its ad hoc and cold storage Hadoop data processing clusters to GCP. To continue the data processing in the cloud, Twitter also had to replicate over 300 PB of data from on-premise HDFS storage systems to Google Cloud Storage (GCS). While working on the data migration, we quickly realized the need for a comprehensive access control framework for the data that we were migrating to the public cloud. Having a robust and transparent access control solution that fades the boundaries between on-premise and public cloud empowers the users and teams to analyze and visualize the data to improve how new features are built on Twitter. \\
This paper showcases Twitter’s approach in designing and implementing a robust identities and access control framework for shared resources in hybrid cloud environments. The framework covers both elements of access control problem space viz. identity life cycle management and access control automation. The paper is structured as follows: Section \ref{sec:background} discusses the related work and background on how identity and access management (IAM) is handled at Twitter and the challenges faced during our migration to the public cloud, section \ref{sec:framework} showcases our identity and access control framework, section \ref{sec:usecase} presents the implementation of our approach in our big data and analytics service at scale, section \ref{sec:future} describes the future work, and finally, we conclude this paper in section \ref{sec:conclusion}.

\section{Background}\label{sec:background}
Extending Twitter’s data processing capabilities from on-premises data centers to the public cloud meant that the data platform was moving from a multi-tenant on-premise-only model to a multi-tenant hybrid cloud model. This means that some part of data processing happens on the clusters on-premises and some of it in the public cloud. Additionally, a multi-tenant cloud infrastructure signifies that we have multiple users co-existing and accessing the same resources on the same system at the same time. Bridging the gap between how identity and access management (IAM) is done on-premise and the IAM solutions offered by GCP is a challenging problem. It is important to have a strong foundation while designing such an IAM framework for multi-tenant platforms to ensure security, privacy, and data protection. While we were working out the design to extend to a hybrid model, we had 3 guiding principles in mind viz., a) Authentication, b) Authorization and c) Auditing, commonly known as the ``AAA” principle \cite{bib8}. Authentication, also known as AuthN, is the process of verifying the identity of a user or a process. We wanted strong authentication methods for all accesses made to data stored at Twitter. Authorization, also known as AuthZ, means having permission to do an act. This includes permissions to read data, write data, create and delete data storage systems and run data processing jobs. We wanted to ensure explicit authorization is granted to access any data stored at Twitter. Auditing means having a log or a trail of actions performed. We wanted the ability to easily determine who performed what actions on the data. Here, Twitter data represents the data stored in an on-premises data center as well as data stored in the public cloud, GCP. Moreover, we also wanted to follow the best practices of access control. Therefore, we designed the framework keeping the Principle of Least Privilege (PLP) central to our efforts. The Principle of Least Privilege asserts that a user should be given only those access privileges that are needed for it to complete the task. Therefore, if a user does not need access to a data storage system or running a data processing job, the user should not have that right \cite{bib9}. Our framework keeps the access control on the shared resources in the cloud as strict as possible to only those identities are have permissions to access corresponding on-premise resources. The least privilege access along with AAA provides a robust way of handling access control in the public cloud. The rest of this section discusses different ways identities are managed on-premises and on the cloud, and forms the basis for the need for a robust framework to bridge the gap between the two.

\subsection{On-premise Identity Management}
Lightweight Directory Access Protocol (LDAP) is an open, industry standard for maintaining a distributed directory information and authentication services \cite{bib10} \cite{bib11}. Using an LDAP-based identity and authentication service is a common norm in the majority of enterprises. When an employee joins Twitter, they get an LDAP account. Each employee who intends to perform data processing also gets a UNIX account to run their ad hoc data processing job. Once the code has been made robust by that user, they move the job to run in a production environment. For production jobs that are scheduled, these services run as a ``headless" UNIX user. Headless users are UNIX system users that are not bound to a particular user and are run without a graphical user interface shell. Headless user accounts that run services are also commonly known as ``service accounts".

\subsection{Google Cloud Identity Management}
Contrary to the on-premise identity management systems, the identity management in GCP primarily happens in one of four ways viz., a) Google Workspace (also known as GSuite) accounts, b) User-managed service accounts, c) default service accounts and d) Google-managed service accounts. When an employee joins Twitter, they also receive a GSuite account. This GSuite account can be easily used with GCP since it is readily integrated into GCP. User-managed service accounts are created by a user and are used to run an application or compute workload like a Virtual Machine. On the other hand, the ``default service accounts" are service accounts that enable a GCP service to deploy jobs that access other GCP resources. Every Google service that runs in GCP has a default service account associated with it. Finally, Google-managed ``system” service accounts are created by Google to access users' resources so that they can act on users' behalf. It is important to note that since GSuite accounts are primarily assigned to human users, the authentication happens via a password. On the other hand, service accounts do not possess passwords. Instead, they have an RSA encryption key pair associated with them for authentication. The scope of this paper is primarily limited to GSuite accounts and user-managed service accounts management. 

\subsection{Security Challenges}
When a service runs on a virtual machine (VM) in GCP, it may need to access data stored in data storage systems, for example, Google Cloud Storage (GCS). This data can be accessed via GSuite account identity or a service account identity. If a human user is running the service on behalf of their credentials then the service will use GSuite account identity to access the data. Otherwise, the service will possess its own service account identity in the cloud. If the service has to be authenticated for data access, then it implies that the credentials for access need to be made available on the VM for that service to use. If GSuite account credentials are made available on the VM, then anyone who has administrative access to that VM can now assume the identity of the GSuite user and then have access to all of the applications supported by GSuite identity, for example, Gmail, Google Drive and so on. That means any system administrator at an organization can now access the emails and documents of any human user using that VM at that organization. Similarly, the administrator can also take over the service account secret keys available on the VM and access sensitive data via that service account. To avoid this security leak, the simpler workaround would be provisioning one admin service account to run all data processing jobs. In this case, there is no clear audit trail because any access to data for various purposes shows up as one service account identity. Additionally, there is no appropriate authorization of data access because all data has to be accessible to this one service account identity. To make it worse, human users can impersonate the service account to run those jobs on their behalf and have access to any data. This practice goes against both the principles of AAA and Least Privilege.

Our framework provides a novel way to have authentication, authorization, and auditing in a multi-tenant hybrid environment of an on-premises and in-cloud setup by providing a 1:1 mapping between LDAP identities and GCP service account identities. Although our framework focuses on implementation in GCP, it can be easily applied to any major cloud computing platform.

\section{Our Framework}\label{sec:framework}
Before we discuss our identity and access control framework for multi-tenant resources in the cloud, it is essential that we visit a few concepts pertaining to GCP. We introduce them in brief as follows:

\begin{itemize}
  \item \textbf{Organization}: An Organization is a root node of the GCP resource hierarchy.
  \item \textbf{Folder}: Folders are nodes that group similar resources like projects, folders, or a combination of both.
  \item \textbf{Project}: Projects form the basis of creating, managing, and enabling all GCP services and APIs.
  \item \textbf{GCS Bucket}: Buckets are basic storage containers that hold your data.
\end{itemize}

Our framework provides a methodology to map on-premise LDAP and UNIX identities to GCP-based service accounts to allow transparent access control across the hybrid environment. Each LDAP account at Twitter, including human and ``headless" accounts, that needs to access any of the data processing resources on GCP are mapped to a corresponding GCP service account. In our paper, we refer to such a service account as the ``mirror" service account, a service account that mirrors the on-premise identity. In the rest of the paper, each mention of the ``user" account includes human users and ``headless" users or service accounts, unless otherwise mentioned.

\begin{figure}[htbp]
\centerline{\includegraphics[width=\linewidth]{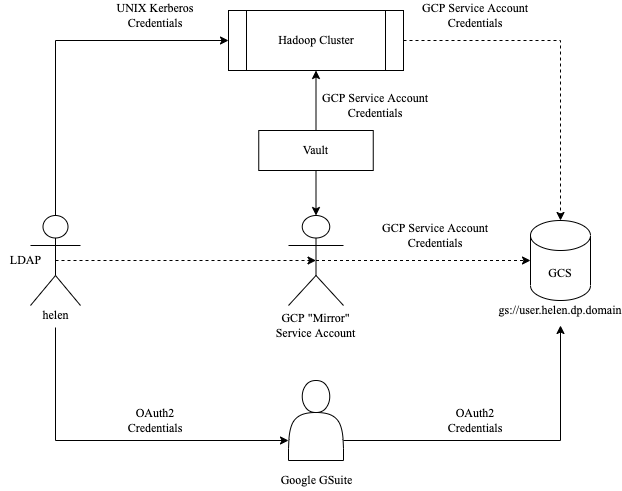}}
\caption{User Authentication in Hybrid Environment}
\label{fig:auth}
\end{figure}

Fig. \ref{fig:auth} showcases the user authentication workflow in our hybrid environment. Each user at Twitter gets an LDAP identity. Users who want to run their data processing jobs programmatically get a UNIX identity. The LDAP and UNIX identities can be mapped 1:1 and may share the same password. For example, a human user with LDAP identity ``helen" possesses the UNIX identity with the same name. Additionally, human users get GSuite identity when they join the organization. While LDAP and UNIX identities can be mapped with each other equally and can be used to run data processing jobs on-premise, GSuite identity is completely separate and it can possess a password that is different from that of LDAP and UNIX identities. LDAP and UNIX identities usually use Kerberos as the authentication protocol. On the other hand, GSuite authentication is managed by Google via an industry-standard OAuth 2.0 protocol. Although LDAP, UNIX, and GSuite identities can be named similarly, there is a gap in interoperability since GSuite identity has its own password. To bridge the gap between on-premise and cloud identities, we introduce a service account called the ``mirror" service account in GCP that is mapped 1:1 to LDAP and UNIX identities. Mirror service accounts are created per-user basis while ensuring that only those mapped identities can act as its mirror service account. A user with LDAP and UNIX identity as ``helen" will get a mirror service account in GCP with the name ``\textit{helen-mirror@service-accounts-project.iam.gserviceaccount.com}". In the next section, we discuss the process to create mirror service accounts identities.

\begin{table}[h]
\caption{Comparing Identities and Authentication Methods}\label{tab:compare-auth}%
\begin{tabular}{lllll}
\toprule
   & LDAP & UNIX   & GSuite   & Service Account  \\
\midrule
Managed by & enterprise    & enterprise   & Google   & Google  \\
AuthN via   & password     & password    & password   & RSA key  \\
Protocol & Kerberos & Kerberos & OAuth 2.0 & OAuth 2.0 \\
\bottomrule
\end{tabular}
\end{table}

\subsection{Process of Mirror Service Accounts Creation}\label{sec:process}
Unlike the on-premise systems where every user gets LDAP and UNIX identities, we wanted to create the mirror identity self-service since a user that wants to use GCP can request their identity in the cloud. Therefore, the process to create the mirror account identities depends on an LDAP group as the source of truth for identities in the cloud. Users must join the LDAP group named ``mirror-account-users" to have their mirror account identity provisioned in the cloud. To make the process of mirror accounts creation automated, we implemented a suite of services. The scope of this paper is limited to one of the functionalities of the Account Creator service that is creating a mirror identity in the cloud. Algorithm \ref{alg:mirror-create} showcases the algorithm in a pseudocode format to create mirror identity in the cloud.

\subsubsection{Preconditioning}
The Account Creator service runs as a cron job based on Cloud Scheduler at a specific interval, for example, 15 minutes. Cloud Scheduler is a fully managed cron job scheduling service offered by GCP. The process starts with verifying that the on-premise LDAP group exists and that it contains members. Once the verification is complete, the process reads the list of users. As mentioned in the sections above, here, users include human users and ``headless" users or LDAP service accounts. For each user, a new service account is created in GCP inside a project named ``service-accounts-project".

\subsubsection{Storing the Identities in the Cloud}
``service-accounts-project" is a special-purpose project that stores all mirror accounts in GCP. This project is created inside a GCP Folder named ``IAMSTORE" to maintain logical separation between identity provisioning and the projects that run data processing jobs as well as shared storage systems like GCS and BigQuery. Since this GCP project is a part of managed service offering, here identity management, no users and mirror accounts that are created as a part of this process are granted any permissions on this project. This is done to ensure that a legitimate mirror service account created in this project cannot create, modify or delete the mirror identities of other users.

\subsubsection{Secret Key Storage}
Unlike human user identities that have passwords, service accounts do not have passwords and they cannot log in via browsers or HTTP cookies. The service accounts also do not belong to the GSuite domain like human user identities. Instead, as discussed in above section, they possess a pair of RSA keys for authentication. Once the mirror service account is created in GCP, the Account Creator service then generates a secret key for the service account in a secret key file. Keeping this secret key file is essential in the multi-tenant environment. The private secret key can be used by an impersonator to request an access token from GCP, assume the identity of the service account and perform unauthorized actions via that service account \cite{bib12}. Additionally, there are a few instances where security keys for a service account can become a security risk. 

\begin{enumerate}
\item Escalation of Privilege: An attacker can gain the access to the secret key and be able to escalate their privilege to access sensitive information.
\item Impersonation: A bad actor can use the service account key to impersonate the service account to carry out malicious activities like reading sensitive personal identifiable information (PII) data.
\item Data Exfiltration: Without secondary platform security arrangements like VPC Service Controls \cite{bib13}, a malicious actor can gain the leaked service account keys to exfiltrate sensitive data outside of the organization.
\end{enumerate}

Therefore, it is essential that the key file can only be accessed by an identity that owns the mirror account in GCP. After the secret key file is generated for a mirror account, the file is stored in a Vault. The Vault is an on-premise secrets store that contains key files for service accounts. Since the GCP infrastructure can be directly connected to on-premise data centers, there is no additional network or proxy connection that needs to be set up for the hybrid cloud environment. This direct connection easily facilitates the communication between GCP and Vault. To ensure that only the LDAP user who owns this mirror service account can access the key file, the Account Creator service makes the LDAP user the owner of the key file. This ensures that no other user can use the secret private key inside this key file and wrongfully impersonate this mirror service account in GCP. Moreover, the user that owns the secret key file for their mirror identity in the cloud does not get the permissions to make changes to the key file. Restricting access to make changes to a secret key file is essential in assuring that the secret keys cannot be manipulated by a bad actor to leak the credentials of a mirror identity. Making an LDAP user the owner of the key file in Vault also assures the 1:1 mapping between LDAP identity on-premises to the mirror identity in the cloud. 

\subsubsection{Acting as a Human Mirror Identity}
Since a human user's GSuite identity is used for authentication and authorization in GCP via a browser, the framework also aids in guaranteeing that human users can programmatically access shared resources in the cloud by creating the human user's mirror identity in the cloud. The process ensures that the correct GSuite human user owns its mirror service account. To achieve this, the Account Creator service applies appropriate permissions for human GSuite users to act as their corresponding mirror service account. After assigning the ``ActAs" permissions, the GSuite identity of the human user can impersonate their mirror service account. However, human users are never permitted to act as a ``headless" user's mirror service account. Since there is no concept of ``headless" users in GSuite, the service only processes human GSuite users for rightful impersonation.

\subsubsection{Key Rotation}
Since encryption keys are long-lived, the probability of a credential breach increases the longer a key is in continuous use. Key rotation mitigates the possibility of such a security breach. When a secret key is rotated, the key is retired and replaced by generating a new key \cite{bib14}. The faster the key is rotated the better are the chances to limit the exposure of the data accessed by that key. Therefore, following the best security practices to ensure that the secret keys are kept secure, the service also rotates the keys after a specific interval of a few days as mandated by the Information Security organization. When the Account Creator service tries to rotate a key, it generates a new key for an existing mirror service account. A new key file is created for the rotated key and stored in the Vault. To avoid data processing outages caused due to key expiration and rotation, the old key file is stored as a valid key for a specific duration and eventually phased out. Our framework also ensures that all versions of the secret key file for a mirror service account are owned by the same on-premise user account that owns the mirror service account in GCP.

\begin{algorithm}
\caption{Mirror Service Account Creation}\label{alg:mirror-create}
\begin{algorithmic}
\Procedure{mirror-account-create}{}
\State System Initialization
\State Verify \textit{mirror-account-users} LDAP Group
\State Read List of Users from \textit{mirror-account-users} 
\For{\textit{each user}}
\State Read LDAP identity
\State Create new a GCP service account in ``service-accounts-project"
\State Generate Key for Mirror Account
\State Store key in Vault
\State Make LDAP identity owner of key
\If{\textit{user.GSuiteAccountExists()}}
\State Assign permission for GSuite identity to act as mirror account
\EndIf
\If{\textit{key.isStale()}}
\State Rotate key
\State Store rotated key in Vault
\EndIf
\EndFor
\EndProcedure
\end{algorithmic}
\end{algorithm}

\subsection{Challenges and Considerations}
While implementing the framework for mirror identities in the cloud, we encountered a few challenges in GCP. These challenges can be categorized into scalability challenges and identity naming issues. This section discusses those challenges and how they can be effectively overcome.

\subsubsection{Scaling the Mirror Identities}
The framework provisions all the mirror service accounts in one central project named ``service-accounts-project" to facilitate easy administration of the identities. A large organization with several hundreds of on-premise LDAP and UNIX identities would need to create hundreds of mirror identities in the cloud. However, there may be limits imposed by the cloud provider on how many service accounts can be created in a project. For example, GCP imposes a default limit of 100 service accounts \cite{bib15}. Therefore, the framework can accommodate only 100 mirror identities in the central project. Since this is not a hard limit, the challenge of scaling your mirror identities can easily be overcome by requesting the cloud platform to increase that limit. Additionally, the number of mirror identities in the cloud can be significantly scaled by making a change in how the identities are created. Instead of storing all the mirror service accounts in a central project, they can be stored across multiple projects based on the organizational unit of on-premise LDAP or UNIX identities. Thus, instead of a central project named ``service-accounts-projects", the mirror service accounts can be stored in different projects like ``dev-service-accounts-project", ``infra-service-accounts-project", ``sales-service-accounts-project" and so on. All these projects can still remain under the same "IAMSTORE" folder for easy administration.

\subsubsection{Naming the Mirror Identities}
Since mirror identities in the cloud are mapped 1:1 to the corresponding LDAP identities, the naming of mirror service accounts are also kept similar. For example, an on-premise LDAP account named ``helen" will get a mirror service account in GCP with the name ``helen-mirror@service-accounts-projectss.iam.gserviceaccount.com". Usually, human LDAP or UNIX accounts do not contain any underscores in their name. However, a ``headless" user may have an underscore character in its name. This means that its corresponding GCP mirror service account must also get an underscore character in the name. The challenge may arise due to an underscore character in the name of on-premise identity because cloud providers like GCP do not allow underscore in the service accounts name. To overcome this challenge, the service framework could have easily replaced the underscore with a hyphen to satisfy the limits imposed by the cloud provider. However, this causes conflicts with on-premise user identities with a hyphen in their name. This would mean that two different on-premise user identities will share the same mirror service account name in the cloud but only one of the users would actually own it. To avoid this conflict, the framework does not provide the mirror service account for an on-premise user identity with an underscore character in it.

\subsection{Following the Best Security Principles}
As discussed in section \ref{sec:background}, our framework follows the best security practices like Authentication, Authorization, and Auditing (AAA) \cite{bib8} and the Principle of Least Privilege \cite{bib9}. 
\subsubsection{AAA Principle}
Our framework enables the individual identities to have their own mirror identities in the cloud to authenticate with. In a multi-tenant environment in the cloud, these identities can easily authenticate their own mirror identities instead of using one admin identity to carry out all data processing jobs. Having unique individual mirror identities in the cloud makes it easier to design authorization techniques. This enables a tightly controlled environment where access control can be more granular. Finally, when a mirror identity is created along with its key file, the Account Creator services log the activities in a central logging sink. Additionally, whenever a user authenticates with their mirror identity and kicks off a data processing job, or reads the data, the activity is logged in the logging sink. This assures the auditing of the actions that take place in the hybrid environment. Finally, when a user account is decommissioned on-premise, its mirror account identity is also decommissioned from the cloud and the key file is invalidated.

\subsubsection{Least Privilege Access}
The framework achieves the principle of least privilege by avoiding the need to have a central administrator service account for running the data processing jobs, and giving access to mirror service account key files to only those identities that are supposed to access them in the cloud. Another benefit of creating a unique mirror identity for an LDAP identity is that the resources in the cloud can be given access to the LDAP identities that are supposed to access specific resources instead of an admin service account. For example, if an admin account ``admin-service-account@dev-team-project.iam.gserviceaccount" inside the project ``dev-team-project" had access to a shared Google Cloud Storage (GCS) bucket ``\textit{gs://production-data}" and if all users in the ``Dev Team" had access to the ``admin-service-account" then that would violate the principle of least privilege since not every identity may require access to the shared resource. Instead, this situation can be overcome by giving access to only those mirror service account identities for the users that require access to the shared resource as a part of the job.

\begin{figure*}[htbp]
\centerline{\includegraphics[width=0.85\textwidth,height=5cm]{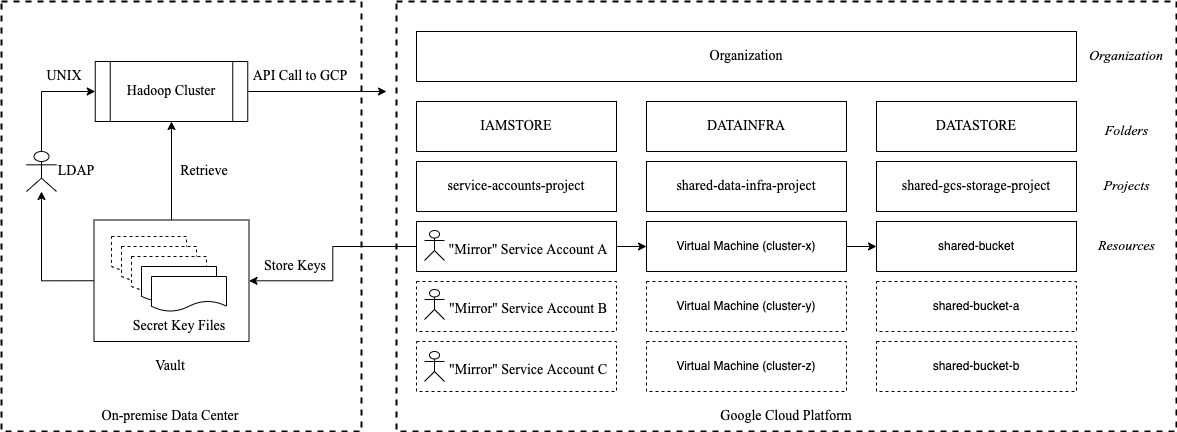}}
\caption{Overview of Identity and Access Management Framework Use Case}
\label{fig:arch}
\end{figure*} 

\section{The Use Case}\label{sec:usecase}
This section showcases the use case of our framework in a multi-tenant data processing environment in a hybrid setup where the data processing clusters are running on-premises and cloud.

\subsection{Project Partly Cloudy}
Before we discuss the use case of our framework in a multi-tenant environment, it is essential to learn about the background and how these multi-tenant data processing clusters work.

Twitter's Hadoop storage systems store more than 300 PB of data across tens of thousands of servers. Every interaction of Twitter users with our services generates log events. These log events are stored in Hadoop File System (HDFS) directories for analytics and eventually, to build better features at Twitter. Additionally, the users who are onboarded to HDFS also get an HDFS directory \cite{bib16}. Here, the users include both - human users and ``headless" users or service accounts. Twitter’s Partly Cloudy is a project to extend data processing at Twitter from an on-premise data center-focused model to a hybrid-cloud model on Google Cloud Platform (GCP). As a part of this project, Twitter migrated its ad-hoc and cold storage Hadoop data processing clusters to GCP and over 300 PB of data from on-premise HDFS storage systems to GCS. Every directory in HDFS for cold storage data processing received a corresponding GCS bucket. For example, ``\textit{/dc1/cluster1/user/helen/some/path/part-001.lzo}" on-premise HDFS directory path is directly mapped to a GCS bucket path named ``\textit{gs://user.helen.dp.domain/some/path/part-001.lzo}". ``helen" here is the human user with an LDAP and UNIX identity. Since data processing in a cloud-native manner was desirable, the ad-hoc Hadoop data processing clusters were also moved to the cloud. These clusters run on virtual machines in a multi-tenant setting.

\subsection{Multi-tenant Data Processing Architecture}
Fig. \ref{fig:arch} showcases the multi-tenant data processing architecture in the hybrid cloud environment. The first part of the architecture is on-premises infrastructure spread across one or more data centers. Here, the data is stored in HDFS directories, and data processing is done via a multitude of Hadoop clusters. The on-premise infrastructure also contains the users with LDAP and UNIX identities. The secrets store called the Vault is set up on-premises to secure it behind the corporate network. As mentioned in section \ref{sec:process}, once the mirror service accounts are created, their secret key files are stored in the Vault. Only on-premise identities can access their corresponding mirror identity key files from the cloud environment. On the other hand, the multi-tenant cloud architecture is divided into at least three parts viz., service account storage, shared data processing jobs, and shared data storage. The mirror service accounts are created inside the project ``service-accounts-project" inside the folder ``IAMSTORE". Multi-tenant data processing jobs are hosted in ``shared-data-infra-project" inside the ``DATAINFRA" folder. The shared data processing jobs run inside an ad-hoc cluster comprising of a large number of virtual machines in the same project. This virtual machine-based cluster mimics the working of the on-premise Hadoop cluster. Just like the on-premise Hadoop cluster, multiple users can submit the data processing jobs in this cluster with their mirror identities. Finally, the data processing jobs read and write shared data that is stored in shared GCS buckets in the project ``shared-gcs-storage-project" inside the ``DATASTORE" project. 

\subsection{Overall Workflow}
When a ``headless" user ``posts-analyze" wants to submit a data processing job in the cluster running in the cloud, it cannot do so with its LDAP or UNIX identity since GCP does not recognize LDAP or UNIX identities. Therefore, it joins the LDAP group that is used as a source of truth for mirror identities in the cloud. The Account Creator service creates a mirror identity for the ``posts-analyze" named ``posts-analyze-mirror@service-accounts-project.iam.gserviceaccount.com". The key file of this service account is generated and stored in the Vault system where only the ``posts-analyze" user owns and can access that key file. Additionally, since ``posts-analyze" has an HDFS directory on-premise, it also receives a corresponding GCS bucket named ``\textit{gs://user.posts-analyze.dp.domain}". Just like the HDFS directory, this bucket is solely owned by ``posts-analyze" by applying appropriate IAM permissions to the ``headless" user's mirror service account. Keeping in mind the AAA principles, since the on-premise HDFS directory is owned by the LDAP identity of the ``posts-analyze" user, its corresponding GCS bucket is owned by its mirror identity in the cloud named ``posts-analyze-mirror". 

\subsubsection{Data Processing on User's Data}
When a data processing job is submitted by ``posts-analyze" user from one of a multitude of data processing clusters like Hadoop clusters in on-premise data centers or cloud, the headless user first initiates the authentication by retrieving the secret key from the Vault and then using the key to authenticate itself. As a part of the data processing job configuration, the user also mentions the path to the GCS bucket that stores the data. If the user mentions the path of the GCS bucket that it owns, then it is automatically authorized to perform read and write actions on the data as per the design of the Partly Cloudy project. However, the user cannot perform read or write actions on the data owned by other users. Finally, the data processing job is started on behalf of the mirror identity of the user. Since the data processing job runs on behalf of mirror identity, the same mirror identity of the user is used to read or write data that is authorized for the mirror identity. Every action that is performed as a part of the data processing jobs is audited and logs are stored in a log sink.

\subsubsection{Reading the Data Owned by Other Users}
There may be a case where the data processing kicked off by a user depends on reading the data owned by other users. Since the access control mechanism for the data storage can be strict and limiting for other users in several use cases, our framework also applies ``reader" groups that provide read-only access to a cloud resource. To gain read-only access to a GCS bucket owned by a different user, the user must be a part of the reader group. To make the user experience homogeneous, the framework creates an on-premise ``reader" LDAP group corresponding to each GCS bucket that a user can join. This LDAP group is then 1:1 mapped to Google group in the cloud via an Onboarder Service. This service was implemented to onboard LDAP identities from LDAP groups to the corresponding Google groups. The design and implementation details of this service are outside the scope of this paper. The corresponding reader Google group for an LDAP reader group in the cloud is given read-only permission on the GCS bucket. To bridge the gap between the LDAP reader group and the Google reader group, the Onboarder Service finds the on-premise identity of the user in the LDAP reader group and adds that mirror identity of that user to the corresponding Google reader group. This way a user that wishes to read the data owned by other users would easily run a data processing job with its mirror identity and use the same mirror identity to perform read-only operations on the data, thereby following the principle of least privilege.

\section{Future Work}\label{sec:future}
The current framework maps the on-premise LDAP identities to mirror account identities in the cloud by provisioning them in one central project named ``service-accounts-projects". As discussed in section \ref{sec:process}, the mirror identities are created in the cloud, only if the user joins the LDAP group ``mirror-account-users". GCP has default limits on how many service accounts can be created in a project. Therefore, to be cognizant of the limit, having the LDAP group as the source of truth puts a check on the number of mirror service accounts that are created in the cloud. However, for a large enterprise, there may be thousands of users that need the mirror identity. In that case, the solution may scale only up to the hard limit of what GCP as a platform can support. Default limits are usually soft limits that can be increased, unlike hard limits. The hard limits are maximum limits imposed by the design and implementation of a product by the cloud provider.

\begin{figure}[htbp]
\centerline{\includegraphics[width=\linewidth]{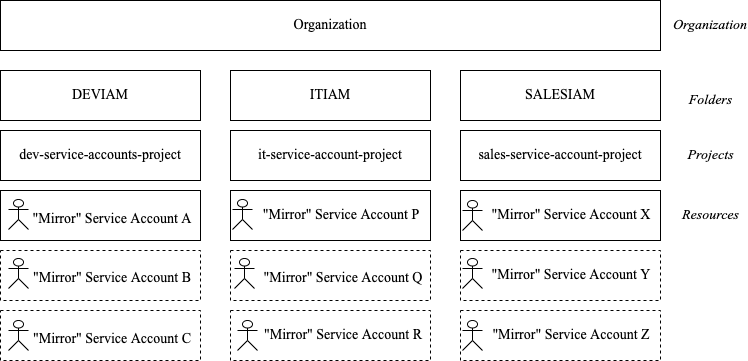}}
\caption{Scaling the Current Framework}
\label{fig:future}
\end{figure}

Therefore, our future work on this paper focuses on scaling the framework to several thousands of mirror identities in the cloud. To scale beyond the default limits of GCP, we propose to divide the project that stores the mirror service accounts into a multitude of projects as shown in Fig. \ref{fig:future}. This division can be based on the functions of different organizations in the enterprise. For example, all developer users' mirror service accounts can be created inside ``dev-service-accounts-projects", all IT user's mirror service accounts can be created inside ``it-service-accounts-projects" and so on. Similarly, although on-premises LDAP groups do not suffer from similar limits in the cloud, they can also be partitioned for the best user experience. Partitioning the identity provisioning among multiple projects can help in alleviating the potential breach of quotas and limits on GCP if the identities were stored in one single project. Since the framework follows the best practices to create a GCP hierarchy in terms of folders and projects, any project that reaches the limit on the number of mirror service accounts can be further partitioned into multiple projects under the same folder. For example, if ``dev-service-accounts-projects" reaches the limit on the number of service accounts, it can further be partitioned into multiple projects while being under the same folder ``DEVIAM" for better administration. Although the framework can be partitioned into multiple projects, the process of provisioning the mirror service accounts, creating the secret key files, storing the key files in the Vault, and assigning the ownership of the key file to its corresponding LDAP user identity remains the same to ensure compliance to the AAA principle.

\section{Conclusion}\label{sec:conclusion}
The paper showcases a novel way of solving the challenges in identity and access control in a multi-tenant environment in hybrid cloud computing by implementing a framework of creating the identities in GCP while ensuring the best security practices. In a multi-tenant environment, multiple users may request access to shared resources. Usually, to avoid complications, many enterprises use a shared admin user account to run data processing jobs. Our framework obviates the need for provisioning and using the shared admin user account by creating identities for individual users in the cloud. Each on-premises user received a mirror service account identity in the cloud and only that user is permitted to use the key file for their mirror identity in the cloud. Moreover, our framework provides more flexibility in providing permissions to specific user mirror identities for reading or writing to shared data resources. The paper also discussed the current implementation of the framework at scale in our data processing production clusters. The future work discusses the ways the framework can be further scaled to accommodate the needs of a large-scale enterprise. The authors hope that the framework will provide the researchers and industry peers with a path to solving identity and access management challenges in a similar multi-tenant hybrid cloud environment.

\section{Acknowledgment}
The work showcased in this paper is the culmination of the hard work of many teams at Twitter as well as Google. The authors would like to thank everybody in Twitter and Google who contributed to designing and implementing this identity and access management framework.

\end{document}